\documentclass[twocolumn]{autart}    % Enable this line and disable the
\usepackage{natbib}
\usepackage{graphicx}          % Include this line if your
\begin{document}

\begin{frontmatter}
\runtitle{Graviton Background from Extra Dimensions}
                          % Running title for regular

\title{A Thermal Graviton Background from Extra Dimensions}%\thanksref{footnoteinfo}}            % Title, preferably not more

%\thanks[footnoteinfo]{This paper was not presented at any IFAC
%meeting. Corresponding author M.~T.~Cicero. Tel. +XXXIX-VI-mmmxxi.
%Fax +XXXIX-VI-mmmxxv.}

\author[Florida]{E.R. Siegel}\ead{siegel@phys.ufl.edu},    % Add the
\author[Florida]{J.N. Fry}%\ead{fry@phys.ufl.edu},               % e-mail address

\address[Florida]{University of Florida, Department of Physics, Gainesville, FL 32611-8440, USA}  % Please supply full addresses here

%I think this uses PACS numbers instead of keywords.

                             % keyword list or with the
                                           % help of the Automatica
                                           % keyword wizard

\begin{abstract}                          % Abstract of not more than 200 words.

Inflationary cosmology predicts a low-amplitude graviton
background across a wide range of frequencies. This Letter shows
that if one or more extra dimensions exist, the graviton
background may have a thermal spectrum instead, dependent on the
fundamental scale of the extra dimensions.  The energy density is
shown to be significant enough that it can affect nucleosynthesis
in a substantial way. The possibility of direct detection of a
thermal graviton background using the 21-cm hydrogen line is
discussed. Alternative explanations for the creation of a thermal
graviton background are also examined.

%The old abstract...
%
%This paper discusses the creation of a thermal graviton background
%due to the presence of one or more extra dimensions.  If extra
%dimensions exist at an energy scale below the reheat temperature
%attained at the end of inflation, gravitational interactions will
%be in equilibrium with other interactions, causing gravitons to
%thermalize. This would result in a relic thermal background of
%gravitons populating the universe today, independent of the type
%of extra dimensions chosen. Other possible explanations capable of
%creating a spectrum of thermal gravitons are shown to conflict
%with observations. Possible methods for direct and indirect
%detection of such a high-frequency relic graviton background are
%also discussed.

%\pacs{98.80.Cq, 98.70.Vc, 95.30.Sf, 04.50.+h}

$PACS:$ 04.50.+h; 95.30.Sf; 98.70.Vc; 98.80.Cq
\end{abstract}

\end{frontmatter}

One of the most powerful windows into the early universe are
backgrounds of particles whose interactions have frozen-out.  The
primordial photon background, the primordial baryon background and
the primordial neutrino background are all examples of particles
that were once in thermal equilibrium.  At various times during
the history of the universe, the interaction rate of the species
in question dropped below the Hubble expansion rate of the
universe, causing the species in question to freeze-out.  The
primordial photon background is observed as the cosmic microwave
background (CMB), the baryon background is observed as stars,
galaxies, and other normal matter, and the neutrino background,
although not yet observed, is a standard component of big bang
cosmology. In addition to these backgrounds, a primordial
background of gravitons (or, equivalently, gravitational waves) is
expected to exist as well, although it, too, has yet to be
detected.  The frequency spectrum and amplitude of this
background have the potential to convey much information about the
early universe. This paper focuses on using the cosmic gravitational
wave background (CGWB) as a probe of extra dimensions.
% in the context of the concordance inflationary cosmology.

%In pre-inflationary standard big-bang cosmology, the hot, dense
%state of the universe is extrapolated back to the Planck scale
%$(m_{pl} \approx 1.22*10^{19} \, \, \rm{GeV})$.  At this scale,
%gravity is of a comparable interaction strength to the other three
%forces.  In this cosmological framework, any particle with a mass
%less than $m_{pl}$ will be created in thermal equilibrium at early
%times. Additionally, due to the energy-dependent strength of the
%gravitational interaction, a thermal spectrum of gravitons will be
%created at $T \sim \mathcal{O}(m_{pl})$ \cite{KT:90}.

The success of the inflationary paradigm \cite{Guth:81} in
resolving many problems associated with the standard big-bang
picture \cite{BBproblems:1} has led to its general acceptance.
Inflationary big bang cosmology predicts a stochastic background
of gravitational waves across all frequencies \cite{Staro:79}, 
\cite{Allen:87}. The amplitude of this background is dependent 
upon the specific model of inflation, but the fractional energy 
density in a stochastic CGWB is constrained \cite{Peebles:04} to be
\begin{equation} \label{eq1}
\Omega_g \leq \mathcal{O}(10^{-10}) \mathrm{.}
\end{equation}

In inflationary cosmology, the predicted CGWB, unlike the CMB and
the neutrino background, is non-thermal.  Gravitational
interactions are not strong enough to produce a thermal CGWB at
temperatures below the Planck scale $(m_{pl} \approx 1.22 \times
10^{19} \, \, \rm{GeV})$.  As the existing particles in the
universe leave the horizon during inflation, the only
major contributions to the energy density will be those particles
created during or after reheating, following the end of inflation.
Unless the reheat temperature $(T_{RH})$ is greater than $m_{pl}$,
gravitational interactions will be too weak to create a thermal
CGWB.  The measurement of the magnitude of the primordial
anisotropies from missions such as COBE/DMR \cite{cobe:1} and WMAP
\cite{wmap:03} provides an upper limit to the energy scale at
which inflation occurs \cite{LL:00}. From this and standard
cosmological arguments \cite{KT:90}, an upper limit on $T_{RH}$
can be derived to be
\begin{equation} \label{eq2}
T_{RH} \simeq  6.7 \times 10^{18} \, \left(g_{*}\right)^{-1/4}
\left(\frac{t_{pl}}{t_{\phi}}\right)^{1/2} \, \, \rm{GeV}
\mathrm{,}
\end{equation}
where $g_{*}$ is the number of relativistic degrees of freedom at
$T_{RH}$, $t_{pl}$ is the Planck time, and $t_\phi$ is the
lifetime of the inflaton.  A stronger upper limit on $T_{RH}$
$(\sim 10^8-10^{10} \, \, \rm{GeV})$ can be obtained from
nucleosynthesis \cite{Sarkar:96} if supersymmetry is assumed.  In
all reasonable cases, however, $T_{RH} \ll m_{pl}$, indicating
that the CGWB is non-thermal in inflationary cosmology.

If the universe contains extra dimensions, however, predictions
about the shape and amplitude of the CGWB may change drastically.
Cosmologies involving extra dimensions have been well-motivated
since Kaluza \cite{Kal:21} and Klein \cite{Klein:26} showed that
classical electromagnetism and general relativity could be unified
in a 5-dimensional framework. More modern scenarios involving
extra dimensions are being explored in particle physics, with most
models possessing either a large volume \cite{ADD:1}, \cite{ADD:2}
or a large curvature \cite{RS:1}, \cite{RS:2}.  Any spatial
dimensions which exist beyond the standard three must be of a
sufficiently small scale that they do not conflict with
gravitational experiments. The 3+1 dimensional gravitational force
law has been verified down to scales of $0.22 \, \, \rm{mm}$
\cite{Eidelman:04}. Thus, if extra dimensions do exist, they must
be smaller than this length scale.\footnote{A possible explanation
for the vast difference in size between the three known spatial
dimensions and any extra dimensions is given in \cite{ChoDet:80}.}
Although there exist many different types of models containing
extra dimensions, there are some general features and signals
common to all of them.

%The general effects of extra dimensions on gravity for a
%4+$\delta$-dimensional universe are as follows.

In the presence of $\delta$ extra spatial dimensions, the
3+$\delta$+1-dimensional action for gravity can be written as
\begin{eqnarray} \label{eq3}
\nonumber\\
\mathcal{S} &=& \int{d^4x \, \left\{\int{d^{\delta}y \sqrt{-g'}
\frac{\mathcal{R'}}{16 \pi G'_N}}+ \sqrt{-g}
\mathcal{L}_{m}\right\}} \mathrm{,} \nonumber\\
%\end{equation}
%\begin{equation} \label{eq6}
 G'_N &=& G_N \frac{m_{pl}^2}{m_{D}^{2+\delta}} \mathrm{,}
\end{eqnarray}
where $g$ is the 4-dimensional metric, $G_N$ is Newton's constant,
$g'$, $G'_N$, and $\mathcal{R'}$ denote the higher-dimensional
counterparts of the metric, Newton's constant, and the Ricci scalar,
respectively, and $m_{D}$ is the fundamental
scale of the higher-dimensional theory. In 3+$\delta$ spatial
dimensions, the strength of the gravitational interactions scale
as $\sim (T/m_{D})^{(1+\delta/2)}$. If $\delta=0$, then
$m_{D}=m_{pl}$, and standard 4-dimensional gravity is recovered.
%If $T_{RH}$ is larger than
%$m_{D}$, a thermal graviton background will be created with relic
%energy density given by equation (\ref{eq1}). %Wait with this - conclusion.

\begin{figure}
\begin{center}
\includegraphics[width=8.4cm]{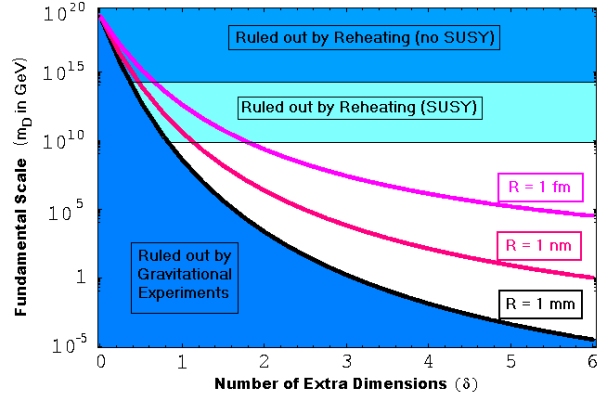}
\caption{Parameter space for the creation of a thermal CGWB in the
context of Large Extra Dimensions. The shaded areas represent
areas ruled out by gravitational experiments and reheating, both
with and without the assumption of supersymmetry.  Certain
assumptions about gravitino physics, as detailed in
\cite{Sarkar:96}, may significantly lower the bound on reheating
with supersymmetry in extra dimensions.} \label{fig1}
\end{center}
\end{figure}

When energies in the universe are higher than the fundamental
scale $m_{D}$, the gravitational coupling strength increases
significantly, as the gravitational field spreads out into the
full spatial volume.  Instead of freezing out at $\sim
\mathcal{O}(m_{pl})$, as in 3+1 dimensions, gravitational
interactions freeze-out at $\sim \mathcal{O}(m_{D})$ \cite{ADD:1}.
($m_{D}$ can be much smaller than $m_{pl}$, and may be as small as
$\sim \rm{TeV}$-scale in some models.) If the gravitational
interactions become strong at an energy scale below the reheat
temperature $(m_D < T_{RH})$, gravitons will have the opportunity
to thermalize, creating a thermal CGWB. Fig. \ref{fig1}
illustrates the available parameter space for the creation of a
thermal CGWB in the case of large extra dimensions, following the
formalism in \cite{Giu:99}.
Other types of extra dimensions have minor quantitative
differences in the shape of their parameter spaces. However, the
qualitative result, the creation of a thermal CGWB if $m_D <
T_{RH}$, is unchanged by the type of extra dimensions chosen.

%Due to large temperatures and densities, interactions were
%stronger and more frequent in the early universe. Interactions
%freeze-out, or fall out of equilibrium, when the interaction rate
%drops below the Hubble expansion rate. Electromagnetic
%interactions freeze-out at a temperature $T \sim 0.2 \, \,
%\rm{eV}$ (CMB decoupling), weak interactions at $T \sim 1 \, \,
%\rm{MeV}$ (neutrino decoupling), and gravitational interactions
%(for standard general relativity) at $T \sim 10^{19} \, \,
%\rm{GeV}$ (graviton decoupling).

%Gauge coupling unification in 4-dimensions is expected to occur at
%$\sim \mathcal{O}(10^{15}-10^{16} \, \, \rm{GeV})$
%\cite{LanPol:93}, with gravitational unification (when
%gravitational interactions become strong) occurring at $\sim
%\mathcal{O}(m_{pl})$. The existence of extra dimensions not only
%lowers the gauge coupling unification scale significantly
%\cite{DDG:98}, but

%A relic background of thermal gravitons
%will be present in the universe today if one or more extra
%dimensions exist at an energy scale below the reheat temperature.
%Calculate from L.E.D. because it's easy.  Maybe it's not important.

%%%%%%%%%%%%%%%%%%%%%%%%%%%%%%%%%%%%%%%%%%%%%%%%%%%%%%%%%%%%%%%%%%%
%%%  Part two -- now that the point has been made, discuss it!  %%%
%%%%%%%%%%%%%%%%%%%%%%%%%%%%%%%%%%%%%%%%%%%%%%%%%%%%%%%%%%%%%%%%%%%

Thus, if extra dimensions do exist, and the fundamental scale of
those dimensions is below the reheat temperature, a relic thermal
CGWB ought to exist today.  Compared to the relic thermal photon
background (the CMB), a thermal CGWB would have the same shape,
statistics, and high degree of isotropy and homogeneity. The
energy density ($\rho_g$) and fractional energy density
($\Omega_g$) of a thermal CGWB are
\begin{equation} \label{eq4}
\rho_g = \frac{\pi^2}{15} \left(\frac{3.91}{g_{*}}\right)^{4/3}
\left(T_{CMB}\right)^4 \mathrm{,}
\end{equation}
\begin{equation} \label{eq5}
\Omega_g \equiv \frac{\rho_g}{\rho_c} \simeq 3.1 \times 10^{-4} \,
({g_{*}})^{- 4/3} \mathrm{,}
\end{equation}
where $\rho_c$ is the critical energy density today, $T_{CMB}$ is
the present temperature of the CMB, and $g_{*}$ is the number of
relativistic degrees of freedom at the scale of $m_{D}$. $g_{*}$
is dependent on the particle content of the universe, i.e. whether
(and at what scale) the universe is supersymmetric, has a KK
tower, etc.  Other quantities, such as the temperature ($T$), peak
frequency ($\nu$), number density ($n$), and entropy density ($s$)
of the thermal CGWB can be derived from the CMB if $g_{*}$ is
known, as
\begin{eqnarray} \label{eq6}
\nonumber\\
n_{g} &=& n_{CMB} \, \left(\frac{3.91}{g_{*}} \right) \mathrm{,}
\qquad s_{g} = s_{CMB} \, \left(\frac{3.91}{g_{*}} \right) \mathrm{,} \nonumber\\
T_{g} &=& T_{CMB} \, \left(\frac{3.91}{g_{*}} \right)^{1/3}
\mathrm{,} \, \, \, \, \nu_{g} = \nu_{CMB}
\left(\frac{3.91}{g_{*}} \right)^{1/3} \mathrm{.}
\end{eqnarray}
These quantities are not dependent on the number of extra
dimensions, as the large discrepancy in size between the three
large spatial dimensions and the $\delta$ extra dimensions
suppresses those corrections by at least a factor of $\sim
10^{-29}$. As an example, if $m_{D}$ is just barely above the
scale of the standard model, then $g_{*} = 106.75$.  The thermal
CGWB then has a temperature of 0.905 Kelvin, a peak frequency of
19 GHz, and a fractional energy density $\Omega_g \simeq 6.1
\times 10^{-7}$.

%Thermal Gravitons.  Other explanations.  No inflation.
%$\dot{G}/G$. PBH decay.

%The only major differences between the photon and graviton
%backgrounds would be in their number densities ($n$), entropy
%densities ($s$), energy densities ($\rho$), and in their peak
%frequencies ($\nu$),
%\begin{eqnarray} \label{eq6}
%\nonumber\\
%n_{g} &=& n_{\gamma} \, \left(\frac{3.91}{g_{*}\left(m_D\right)}
%\right) \mathrm{,}
%\qquad s_{g} = s_{\gamma} \, \left(\frac{3.91}{g_{*}\left(m_D\right)} \right) \mathrm{,} \nonumber\\
%\rho_{g} &=& \rho_{\gamma} \,
%\left(\frac{3.91}{g_{*}\left(m_D\right)} \right)^{4/3} \mathrm{,}
%\, \, \, \, \nu_{g} = \nu_{\gamma}
%\left(\frac{3.91}{g_{*}\left(m_D\right)} \right)^{1/3} \mathrm{.}
%\end{eqnarray}
%All of the differences between the CMB and the thermal CGWB are
%accounted for by a single parameter, $g_{*}\left(m_D\right)$, or
%the effective number of relativistic degrees of freedom at
%gravitational freeze-out.   For $m_D$ just above the scale of the
%standard model $(g_{*}=106.75)$, this would create a graviton
%background peaked at a frequency of $\sim 19 \, \, \rm{GHz}$, with
%a fractional energy density of $\Omega_g \simeq 6.1 * 10^{-7}$.

%Gravitational wave detectors -- signals.

Although the fractional graviton energy density is expected to be
small today, it may be detectable either indirectly or directly.
Nucleosynthesis provides an indirect testing ground for a thermal
CGWB. Standard big-bang nucleosynthesis predicts a helium-4
abundance of $Y_p = 0.2481 \pm 0.0004$ \cite{ManSer:04}.  With a
thermal CGWB included, the expansion rate of the universe is
slightly increased, causing neutron-proton interconversion to
freeze-out slightly earlier. A thermal CGWB can be effectively
parameterized as neutrinos, as they serve the same function at
that epoch in the universe (as non-collisional radiation).  The
effective number of neutrino species is increased by $N_{\nu -
eff} \simeq 27.1 \, (g_*)^{-4/3}$, or $\simeq 0.054$ (for
$g_* = 106.75$). This would yield a new prediction of $Y_p =
0.2489 \pm 0.0004$ for helium-4.  Although observations are not
yet able to discriminate between these two values, the constraints
are tightening with the advent of recent data \cite{Cyburt:04}. An
increase in the precision of various measurements, along with an
improvement in the systematic uncertainties, may allow for the
indirect detection of a thermal CGWB.

Direct detection of a thermal CGWB is much more challenging, but
would provide quite strong evidence for its existence.
Conventional gravitational-wave detectors include cryogenic
resonant detectors \cite{Frossati:03}, which have evolved from the
bars of Weber \cite{Weber:60}, doppler spacecraft tracking, and
laser interferometers \cite{Thorne:80}.  The maximum frequency
that these detectors can probe lies in the kHz regime, whereas a
thermal CGWB requires GHz-range detectors.  An interesting
possibility for detection may lie in the broadening of quantum
emission lines due to a thermal CGWB.  Individual photons
experience a frequency shift due to gravitational waves
\cite{EstWahl:75}. For a large sample of radio-frequency photons
in a gravitational wave background, the observed line width $(W)$
will broaden by
\begin{equation} \label{eq7}
\Delta W \sim h_0 \sim \frac{\sqrt{\Omega_g}}{\nu \, t_0} \sim
10^{-31} \left( \frac{106.75}{g_{*}} \right)^{1/3} \mathrm{,}
\end{equation}
where $t_0$ is the present age of the universe, $\nu$ is the peak
frequency of the thermal CGWB and $h_0$ is the metric perturbation
today due to the thermal CGWB \cite{Carr:80}.
%The numbers chosen are again
%typical of $m_D$ just above the energy scale of the standard
%model.
As $\mathcal{O}(10^{-31})$ is a very small broadening, a radio
line with a narrow natural width is the preferred candidate to
observe this effect.  One possibility for this type of observation
is the $21$-cm emission line of atomic hydrogen.  So long as the
emitting atoms and the detectors are sufficiently cooled,
broadening due to thermal noise will be suppressed below $\Delta
W$.
%
%The stuff that's below is based on the incorrect assumption that
%the gravitational broadening is path-length dependent.
%
%A radio line with a very narrow natural width is necessary, as
%$\Delta W$ ought to dominate $W$. Additionally, thermal broadening
%due to motion of the emitting atoms must be suppressed below
%$\Delta W$. One possible candidate experiment is to cryogenically
%cool individual hydrogen atoms, and to measure the width of the
%$21$-cm emission line.
%
Because the lifetime $(1 / \Gamma)$ of the excited state of
hydrogen is large $(\sim 10^7 \, \, \rm{yr})$ and the frequency of
the emitted light $(\nu_\gamma)$ is high $(\sim 10^9 \, \,
\rm{Hz})$, the natural width $(W)$ is among the smallest known
\begin{equation} \label{eq8}
W = \frac{\Gamma}{\nu_\gamma} \simeq \frac{2.869 \times 10^{-15}
\, \, s^{-1}}{1.42040575179 \times 10^{9} \, \, s^{-1}} \simeq
2.02 \times 10^{-24} \mathrm{.}
\end{equation}
%
%The stuff that's below is based on the incorrect assumption that
%the gravitational broadening is path-length dependent.
%
%For $m_D$ just above the scale of the standard model, a path
%length $d$ of $\sim 10^{15}$ cm is necessary to increase $\Delta
%W$ above $W$. At this scale, the individual atoms must be
%cryogenically cooled to below $\simeq 22 \, \, \mathrm{pK}$ to
%suppress thermal broadening below $\Delta W$. The experiment could
%be done at higher temperatures, but would require a prohibitively
%long path length. As an example, a temperature of $1 \, \, \mu
%\mathrm{K}$ requires a path length of $\sim 10^{23}$ cm, which is
%far beyond the limits of a reasonable experiment.
%
The width of the $21$-cm line is regrettably seven orders of
magnitude larger than the expected broadening due to a thermal
CGWB. Extraordinarily accurate measurements would need to be taken
for direct detection of this background. Additionally,
temperatures of the atoms and detectors would need to be
cryogenically cooled to $\sim 10^{-18}$ Kelvin to suppress thermal
noise below $\Delta W$. This last criterion is far beyond the
reach of current technology, and either a major advance or
experimental innovation would be required to measure the desired
effect using this technique.

%%%%%%%%%%%%%%%%%%%%%%%%%%%%%%%%%%%%%%%%%%%%%%%%%%%%%%%%%%%%%%%%
%%%  Alternative Possibilities for a thermal CGWB            %%%
%%%%%%%%%%%%%%%%%%%%%%%%%%%%%%%%%%%%%%%%%%%%%%%%%%%%%%%%%%%%%%%%

Extra dimensions are not the only possible explanation for the
existence of a thermal CGWB.  Currently, there are three known
alternative explanations that would also create a thermal CGWB.
They are as follows: there was no inflation, there was a spectrum
of low-mass primordial black holes that have decayed by the
present epoch, or the gravitational constant is time-varying
(the Dirac hypothesis). Each alternative is shown below to face
difficulties that may make extra dimensions an attractive
explanation for the creation of a thermal CGWB.

The predictions of inflation are numerous \cite{LL:00}, and many
have been successfully confirmed by WMAP \cite{wmap:03}. The major
successes of inflation include providing explanations for the
observed homogeneity, isotropy, flatness, absence of magnetic
monopoles, and origin of anisotropies in the universe.
Additionally, confirmed predictions include a scale-invariant
matter power spectrum, an $\Omega=1$ universe, and the spectrum of
CMB anisotropies. To explain a thermal CGWB by eliminating
inflation would require alternative explanations for each of the
predictions above. Although alternative theories have been
proposed, as in \cite{HW:02}, they have been shown to face
significant difficulties \cite{KLM:02}. The successes of inflation
appear to suggest that it may likely provide an accurate
description of the early universe.

Primordial black holes with masses less than $10^{15} \, \,
\rm{g}$ would have decayed by today, producing thermal photons,
gravitons, and other forms of radiation.  Density fluctuations in
the early universe, in order to produce a large mass fraction of
low-mass primordial black holes, and not to produce too large of a
mass fraction of high-mass ones, favor a spectral index $n$ that
is less than or equal to $2/3$ \cite{Carr:76}. Accepting the
observed scale-invariant $(n \simeq 1)$ spectrum of density
fluctuations \cite{spectral:04} may disfavor primordial black
holes as a reasonable candidate for creating a thermal CGWB.

The Dirac hypothesis states that the difference in magnitude
between the gravitational and electromagnetic coupling strengths
arises due to time evolution of the couplings \cite{Dirac:37}. If
true, gravitational coupling would have been stronger in the early
universe. At temperatures well below the Planck scale, gravity
would have been unified with the other forces, creating a thermal
CGWB at that epoch. However, this hypothesis produces consequences
for cosmological models that are difficult to reconcile
\cite{Steigman:78}, and any time variation is severely constrained
by geophysical and astronomical observations \cite{SisVuc:90}. The
acceptable limits for variation are small enough that they cannot
increase coupling sufficiently to generate a thermal CGWB
subsequent to the end of inflation. The difficulties faced by each
of these alternative explanations points towards extra dimensions
as perhaps the leading candidate for the creation of a thermal
CGWB.

%Alternatives ruled out -- inform about problems with building a
%complete model (moduli/inflation) -- then write concluding paragraph.

%Although evidence for the existence of a thermal CGWB would
%provide strong support for extra dimensions with $m_D < T_{RH}$,
There exist two major obstacles to the construction of a more
complete phenomenological model containing extra dimensions with
$m_D < T_{RH}$. The first of these is the moduli problem
\cite{Randall:95}.  String moduli interactions with standard model
fields are highly suppressed, leading to a long lifetime of the
string moduli.  String moduli decay, however, must be consistent
with astrophysical constraints \cite{Ellis:92}. To accomplish
this, string moduli need either a small production amplitude or
very specific decay channels, which both require fine-tuning.
%By lowering the fundamental scale significantly
%$(m_D \ll m_{pl})$, this problem is exacerbated.
%
%Possible solutions, such as in \cite{Lyth:95},
%exist for this problem at $m_{pl}$, but not presently at $m_D$.
The second problem is the overproduction of long-wavelength tensor
modes from inflation \cite{Chung:00}, \cite{Giudice:02}. While the
short-wavelength modes (the modes inside the horizon when
gravitational interactions freeze-out) will thermalize,
gravitational waves of longer wavelengths will be unaffected.  As
the scale of inflation must be above $m_D$, the amplitude of these
waves is expected to be large.  This would leave an unacceptable
imprint in the CMB. Both problems arise from the fact that at
energies above $m_D$, macroscopic gravity breaks down
\cite{Hebecker:01}. Although these problems may not be resolved
until a quantum theory of gravity is realized, they do not change
the fact that a thermal CGWB would arise from extra dimensions
with $m_D < T_{RH}$.

%The construction of a complete model of the universe above the
%fundamental scale, as shown here, is not without its difficulties.
%In spite of these problems and uncertainties, the detection of a
%thermal CGWB would provide strong evidence for extra dimensions
%with $m_D < T_{RH}$, and should stimulate the search for the
%resolution of these problems.

This work has attempted to show that extra dimensions may be
responsible for the production of a thermal gravitational wave
background. A thermal CGWB, as opposed to the stochastic CGWB of
standard inflationary cosmology, is a prediction of extra
dimensions with a scale below the reheat temperature. The
detection of a thermal CGWB, although challenging at present,
would provide strong evidence for the existence of extra
dimensions. The detected absence of a thermal CGWB would
conversely disfavor the existence of extra dimensions up to the
energy scale of the reheat temperature.

%\section{Introduction}

%\begin{figure}
%\begin{center}
%\includegraphics[height=4cm]{jcaesar.eps}    % The printed column
%\caption{Gaius Julius Caesar, 100--44 B.C.}  % width is 8.4 cm.
%\label{fig1}                                 % Size the figures
%\end{center}                                 % accordingly.
%\end{figure}

% OR

%\begin{figure}
%\begin{center}
%\epsfig{file=jcaesar,width=7cm}
%\caption{Gaius Julius Caesar, 100--44 B.C.}
%\label{fig1}
%\end{center}
%\end{figure}

%\subsection{A subsection}
%Marcus Tullius Cicero, 106--43 B.C. was a Roman statesman, orator,
%and philosopher.  A major figure in the last years of the Republic,
%he is best known for his orations against Catiline\footnote{
%This footnote should be very brief.}
%and for his mastery of Latin prose \cite{Heritage:92}. He was a
%contemporary of Julius Caesar (Fig.~\ref{fig1}).

%\section{The argument}
%Some words might be appropriate describing equation~(\ref{e1}), if
%we had but time and space enough.
%\begin{equation} \label{e1}
%{{\partial F}\over {\partial t}} =
%D{{\partial^2 F}\over {\partial x^2}}.
%\end{equation}
%See \cite{Abl:56}, \cite{AbTaRu:54}, \cite{Keo:58} and
%\cite{Pow:85}.
%This equation goes far beyond the celebrated theorem ascribed to the great
%Pythagoras by his followers.
%\begin{thm}
%The square of the length of the hypotenuse of a right triangle equals the sum of the squares
%of the lengths of the other two sides.
%\end{thm}
%\section{Epilogue}
%A word or two to conclude, and this even includes some inline
%maths:  $R(x,t)\sim t^{-\beta}g(x/t^\alpha)\exp(-|x|/t^\alpha)$.

\begin{ack}                               % Place acknowledgements here.
We acknowledge Dan Chung, Steve Detweiler, Lisa Everett, \`{E}anna
Flanagan, Konstantin Matchev, Pierre Ramond and Bernard Whiting
for their helpful discussions.  ERS acknowledges the University of
Florida's Alumni Fellowship for funding.
\end{ack}

%\bibliographystyle{unsrt}        % Include this if you use bibtex
%\bibliography{draft}           % and a bib file to produce the
                                 % bibliography (preferred). The
                                 % correct style is generated by
                                 % Elsevier at the time of printing.

%\appendix
%\section{A summary of Latin grammar}    % Each appendix must have a short title.
%\section{Some Latin vocabulary}         % Sections and subsections are supported
                                        % in the appendices.
\end{document}